\documentclass[twocolumn,showpacs,prl,amsmath,amssymb,superscriptaddress]{revtex4}

\usepackage{graphicx}
\usepackage{dcolumn}
\usepackage{bm}
\usepackage{mathrsfs}

\begin{document}

\title{Electron spin resonance on a 2-dimensional electron gas in a single AlAs quantum well}

\author{M.~Schulte}
\affiliation{2. Physikalisches Institut Universit{\"a}t Stuttgart,
Pfaffenwaldring 57, 70569 Stuttgart, Germany}
\author{J.~G.~S.~Lok}
\affiliation{Max-Planck-Institut f\"ur Festk\"orperforschung, Heisenbergstrasse 1,
70569 Stuttgart, Germany}
\author{G. Denninger}
\affiliation{2. Physikalisches Institut Universit{\"a}t Stuttgart,
Pfaffenwaldring 57, 70569 Stuttgart, Germany}
\author{W.~Dietsche}
\affiliation{Max-Planck-Institut f\"ur Festk\"orperforschung, Heisenbergstrasse 1,
70569 Stuttgart, Germany}

\date{July 18, 2004}

\begin{abstract}
Direct electron spin resonance (ESR) on a high mobility two dimensional electron gas in a
single AlAs quantum well reveals an electronic $g$-factor of 1.991 at 9.35 GHz and 1.989 at
34 GHz with a minimum linewidth of 7 Gauss. The ESR amplitude and its temperature
dependence suggest that the signal originates from the effective magnetic field caused by
the spin orbit-interaction and a modulation of the electron wavevector caused
by the microwave electric field. This contrasts markedly to conventional ESR that detects
through the microwave magnetic field.

\end{abstract}

\pacs{76.30.Pk, 73.21.Fg, 71.70.Ej}

\maketitle

\par
ESR has long been used to extract $g$-factors and $g$-factor anisotropies of
different kinds of solids and molecules, thus providing experimental verification
for bandstructure calculations in solids and structure calculations in molecules.
Additionally spin-lattice relaxation times ($T_1$) and spin-spin dephasing
times ($T_2$) can be determined~\cite{Poole}. More recently ESR
has successfully been employed to study $g$-factors and spin relaxation of
2D electrons in Si/SiC~\cite{Nestle} and Si/SiGe~\cite{Jantsch1} structures.
ESR on 2D systems also provided information about 2D electron-donor exchange
tunnelling~\cite{Kummerer} and on potential fluctuations caused by remote
doping~\cite{Jantsch2,Wilamowski1}, without the need for Ohmic contacts to the samples.
Moreover, from the dependence of the $g$-factor anisotropy on Fermi wavevector and from
the dependence of the $g$-factor on angle between microwave field and static magnetic
field, recently the (tiny) Bychkov-Rashba spin-orbit interaction of 2D electrons in Si/SiGe
samples could be determined~\cite{Wilamowski2,Wilamowski3}. In this paper we show that in
high mobility 2D samples, this spin-orbit interaction allows to resonantly manipulate the
electron spin by means of GHz {\em electric} fields.
\par
Direct ESR on a two dimensional electron gas (2DEG) has proved difficult because of the typically
small number of spins in the 2DEG. So far it has been restricted to Si (either in Si/SiC or in
Si/SiGe samples) because of its favourable physical properties. As the sensitivity of ESR
is proportional to the inverse of the linewidth squared, narrow linewidths are a prerequisite.
In Si linewidths down to 3 $\mu T$ are observed~\cite{Wilamowski3}, as little $T_1$-broadening
occurs. This is because Si has a rather small spin-orbit (SO) interaction. Also it has only one
isotope with nuclear spin ($^{29}$Si) which additionally has only a small natural abundance
(4.7 \%).  This contrasts markedly to the III-V semiconductors where there are many
isotopes with nuclear spin ($^{69}$Ga, $^{71}$Ga, $^{27}$Al, $^{75}$As, $^{115}$In, $^{31}$P etc.)
with large natural abundance, many of which have a strong SO coupling. This leads to
considerable line broadening and at low temperatures, where ESR usually has the best
sensitivity, to large hyperfine fields that vary slowly with time. Consequently direct
ESR has never been demonstrated on 2D electrons in III-V semiconductors.
\par
Here, we present the first direct ESR on a 2DEG in a III-V semiconductor. We study ESR of
high mobility 2D electrons in a single AlAs quantum well. At 9.35 GHz and at 34
GHz $g$-factors of 1.991 and 1.989 were determined respectively.
By rotating the sample in the cavity we demonstrate
that our ESR originates from the microwave {\em electric} field ($E_1$-field) and not
from the microwave magnetic field ($B_1$-field). For small power ($P$) of the $E_1$-field,
the ESR follows a $P^{0.5}$-law, but for larger powers, the exponent increases
to $\sim$1. The temperature dependence of the ESR is much stronger than the
2D magnetisation expected for such a system~\cite{Nestle}.
Our observations can be explained by assuming that the spin transitions occur through the
effective magnetic field caused by SO interaction and the modulation of the electron wavevector
around $k_F$ induced by the microwave $E_1$-field.
\par
Our sample is a 2$\times$4 mm$^2$ piece of a MBE-overgrown GaAs wafer that contains a 15
nm wide AlAs quantum well (QW) flanked by Al$_{0.45}$Ga$_{0.55}$As barriers. It is volume
doped with Si (4$\times$ 10$^{18}$ cm$^{-3}$) over 30 nm with spacers of 40 and 30 nm below
and above the QW. Transport measurements
on samples from the same wafer,
reveal that the 2D
electrons occupy both in-plane X-valleys~\cite{Lok}. The carrier density is 2.5 $\cdot 10^{15}$
m$^{-2}$ and at 4~K the mobility is 12.5 m$^{2}$/Vs, which compares well with the best results
found in literature~\cite{Poortere2}. The GaAs substrate of our sample was thinned to
$\approx$80~$\mu$m, to minimise dielectric disturbances of the cavity modes ($\epsilon_{GaAs}$=13.1).
\par
In an ESR experiment, the absorption of microwaves by magnetic dipole transitions in a
sample is measured. The (static) magnetic field ($B_0$) is swept at fixed microwave
frequency ($f$) and the reflected microwave power ($P$) is detected. A feature is observed at
the resonant condition, given by $g\mu_BB_0$=$hf$, with $g$ the $g$-factor of the material.
In our experiment (as in most others), $B_0$ is slightly modulated and
ESR ($\propto \sqrt{P}$) is lock-in detected.
\par
\begin{figure}[t]
\centering
\includegraphics[width=85mm]{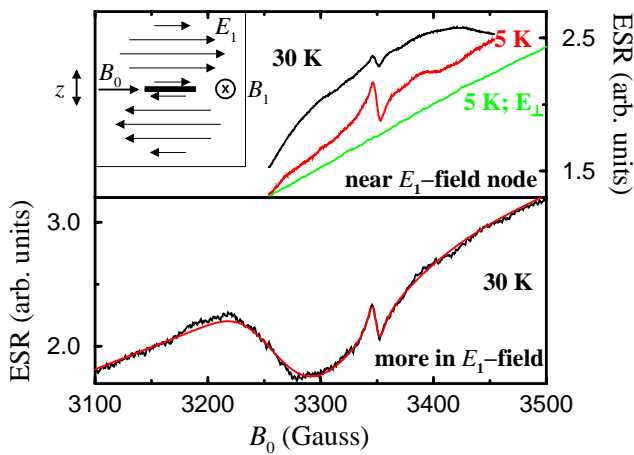}
\caption{Top: ESR at 30~K and 5~K measured at 9.35 GHz and 20 mW with the sample near a node
in the $E_1$-field. The second trace at 5~K demonstrates the absence of ESR when the QW is
oriented perpendicular to the $E_1$-field. The inset shows the orientation of the QW (solid
black bar) with respect to the static and microwave fields for the top 2 traces. Bottom: ESR
with the sample in more $E_1$-field, i.e. shifted by 1 mm ($\approx$1/30 $\lambda$) in the
$z$-direction compared to the inset.}
\label{xbandsignal1}
\end{figure}

Fig.~\ref{xbandsignal1} plots the ESR versus magnetic field at 9.35 GHz (X-band), measured
with a Bruker spectrometer with automatic frequency control (AFC) in a rectangular cavity
(TE$_{102}$ mode, Bruker ER410ST). Data in the top panel are measured with the sample positioned
near the node in the $E_1$-field. The top trace was measured at 30~K with the QW parallel to
the $B_1$- and $E_1$-fields and clearly shows a feature near 3349 Gauss. Upon decreasing the
temperature to 5~K (middle trace), the feature grows in amplitude, but stays at the same position.
The data at 5~K is reasonably well fitted by the derivate of a single Lorentzian.
Because of the AFC only the absorptive part of the signal can be measured.
The line has a width of 7 Gauss and is centred around 3349 Gauss, which yields a $g$-factor
of 1.991. We note that for temperatures above $\sim$10~K a single Lorentzian gives a less
satisfactory fit to the data, suggesting that there is a (small) additional contribution to
the signal~\cite{foot1}.
\par
The observation of a line with a width of 7 Gauss and a signal to noise ratio of 10
represents a puzzle, as the sensitivity of the cavity at a power of 20 mW is only
3$\times$10$^{10}$ spins for a linewidth of 1 Gauss. As our sample contains only
2$\times$10$^{10}$ 2D electrons, for a linewidth of 7 Gauss the ESR should be 74
times smaller than the noise level. We note that the electrons at the Si dopants
in the barrier material (Al$_{0.45}$Ga$_{0.55}$As) have a very different $g$-factor
(+0.6~\cite{Weisbuch}) and cannot account for the signal. Nonetheless, to verify that
the signal originates from the 2D system, ESR was also measured with the QW perpendicular to
the $E_1$-field (bottom curve in the top of fig.~\ref{xbandsignal1}). In such orientation
no ESR was observed, proving that the signal does not originates from 3D (doping) layers.
Moreover, it further proves that the measured ESR does not originate from the $B_1$-field as
is normally the
case~\cite{Nestle,Jantsch1,Kummerer,Jantsch2,Wilamowski1,Wilamowski2,Wilamowski3,WilamowskiPRB2004}
(note that the $B_1$-field is still parallel to the QW and perpendicular to $B_0$),
but surprisingly, that it is caused by an in-plane $E_1$-field.
\par
Fig.~\ref{xbandsignal1} bottom, plots the ESR at 30~K with the sample placed more in the $E_1$-field
in order to enhance the signal. Indeed, the ESR-feature can be more clearly seen, but it is
superimposed on a rather large, broad and very asymmetric background. This background is well
fitted (solid line) with a statistic extreme value function
($I$=exp(1-$z$-exp(-$z$)); $z$=($x$-$x_c$)/$w$ with $x_c$ the centre position (3302 Gauss) and
$w$ the width of the distribution (48 Gauss)). Such a function is commonly used in non-linear
physics to describe the distribution of first moments of underlying
distributions~\cite{extremevalue}, but at present its physical meaning is unclear. Naively
thinking, the extent of the fit from 3200-3450 Gauss implies that a whole spread of $g$-factors
is contributing, possibly originating from electrons that are accelerated high up in the
conduction band.

\begin{figure}[t]
\centering
\includegraphics[width=85mm]{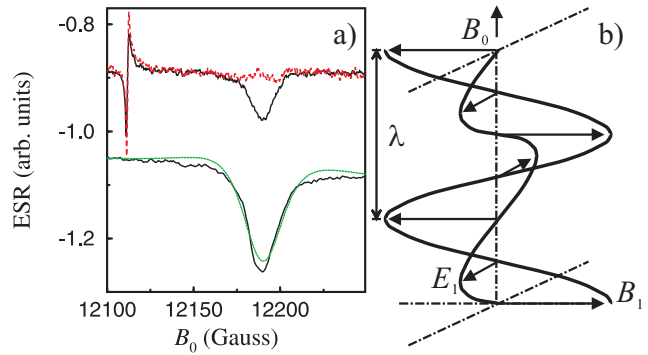}
\caption{a) Top: ESR at 30 K measured at 33.94 GHz and 10 mW with the sample and a Li:LiF
marker ($g=2.002293$) present (solid line) and with Li:LiF only (dashed line).
Bottom: ESR at 22K (solid line) with a dispersive single Lorentzian fit (dotted line). 
b) Orientation of the $B_0$, $B_1$ and $E_1$-fields. The QW is perpendicular to $B_0$ and
$\approx$1/4 $\lambda$ above the bottom of the cavity (near the $E_1$-field maximum).
}
\label{qbandsignal}
\end{figure}

\par
Fig.~\ref{qbandsignal} plots the ESR at 33.94 GHz (Q-band) measured in an oversized, home-built
Fabry-Perot cavity without AFC. In this cavity, $B_0$ is perpendicular to both the $E_1$- and
$B_1$-fields. The sample is mounted perpendicular to $B_0$ and close to the $E_1$-field maximum.
Top traces plot the ESR of the sample and a Li:LiF marker (solid line), and that of the Li:LiF
marker only (dashed line) at 30~K. The marker was included to determine the precise $g$-factor
of the 2D electrons. It further allows to determine the phase of the ESR of the 2D electrons.
The bottom trace (solid line) plots the ESR at 22~K. In all ESR traces with the sample in the
cavity a clear feature at 12190 Gauss appears, which corresponds to a $g$-factor of 1.989. In
the temperature range studied (5-35~K), this $g$-factor is constant. The most striking feature
however, is that the ESR of the 2D electrons is mostly dispersive, whereas that of the marker
is (as it should be) almost completely absorptive. The ESR of the 2D electrons is thus significantly
phase delayed compared to the normal $B_1$-field induced transitions in the marker. The dotted
line is a single dispersive Lorentzian fit and describes the largest part of the ESR-signal.
Just as for 2D electrons in Si~\cite{WilamowskiPRB2004} there is a small additional contribution
to the ESR signal~\cite{foot1}.
\par
Before presenting the temperature and power dependencies of the ESR, we first comment upon its
possible origin. From the absence of any ESR when the QW is perpendicular to the $E_1$-field,
we conclude that it arises from an in-plane $E_1$-field. This $E_1$-field can in principle
cause ESR in two ways. First, it accelerates the high mobility electrons, thus inducing
currents in the sample, which in turn generate a magnetic field ($B_2$). The component of
$B_2$ perpendicular to the $B_0$-field can cause transitions, resulting in an ESR-signal.
In X-band with $P$=20 mW, $\lvert E_1 \rvert _{max}$ is 240 V/m. The measured sheet
conductivity of the 2DEG at 4~K is 5.0 $\times$ 10$^{-3}$ ($\Omega ^{-1}$m$^{-1}$). If
we mimic the QW as an infinite thin metal plate, then the $B_2$-field ($\equiv$ 1/2 $\mu_0 j$)
is 8 mGauss, more than one order of magnitude {\em lower} than the $B_1$-field (0.2~Gauss).
\par
Second, we note that for high enough mobility samples, the $E_1$-field accelerates /
decelerates the electrons periodically in each half of the microwave cycle. This leads to a
modulation of the electron wavevector, which in materials with spin-orbit (SO) interaction
will cause an effective magnetic field that acts on the electron spins only.  In the X-band
cavity only the effective magnetic field due to the Bychkov-Rashba~\cite{Rashba,Bychkov} part
of the SO interaction is perpendicular to $B_0$ and can cause transitions. For the orientation
in Q-band, also the Dresselhaus~\cite{Dresselhaus} part contributes to the effective magnetic
field. Because the AlAs crystal structure lacks inversion symmetry and because the QW-structure
is not symmetric in the growth direction, both parts of the SO interaction should be present.
A necessary condition for the appearance of a wavevector modulation is that the scattering
time of the electrons ($\tau$) is comparable to or larger than the inverse microwave
frequency so that at least part of the electrons follow a cycle of the $E_1$-field
without being scattered. This is indeed the case in our samples; from the mobility
at 4~K (12.5 m$^2$/Vs), using an effective mass of 0.46 m$_e$, we obtain a scattering
time of 33 ps which is comparable to the 100 ps (X-band) or 30 ps (Q-band) inverse microwave
frequency. 
\par
We now estimate the magnitude of the effective magnetic field.
The Bychkov-Rashba SO interaction for a sample that is not symmetric in the growth direction
($z$) can be written as $\mathscr{H}_{BR} = \alpha_{BR}({\bf k} \times \sigma) \cdot {\bf
e}_z$~\cite{Rashba,Bychkov}, with $\alpha_{BR}$ the Bychkov-Rashba constant of the material,
{\bf k} the electron wavevector and $\sigma$ the Pauli spin matrices. It is now evident that
the spin and orbital motion are coupled. To translate the energy into an effective magnetic
field, we divide by 1/2 $g
\mu_B$. This gives an effective field of {\bf B}$_{BR} = (2 \alpha_{BR}/g \mu_B)${\bf k}
$\times${\bf e}$_z$. For high enough mobilities, the $E_1$-field modulates the
wavevector of the electrons, i.e. {\bf k} = {\bf k}$_F$ + $\Delta$ {\bf k}. For X-band
assuming an infinite scattering time, $\Delta$ {\bf k} becomes e$\lvert E_1 \rvert$/(h$f$).
Since the scattering time is not infinite but somewhat smaller than the inverse microwave
frequency, and to estimate $\Delta$ {\bf k} in Q-band and to determine the phase of the
effective magnetic field with respect to the $B_1$-field, we performed a Monte
Carlo simulation, assuming an isotropic effective mass and an energy independent scattering
time taken from transport experiments. We integrate the force on the electron
(-$eE_1$ for X-band and -$e$($E_1$+$v\times B_0$) for Q-band) and after a fixed time of
flight ($\ll \tau$), we determine whether the electron scatters or not. If it scatters, it
restarts at the Fermi energy. In X-band with
$P$=20 mW, $\lvert E_1 \rvert_{max}$ is $\sim$240 V/m and $\Delta$ {\bf k} becomes 2$\times
10^6$ m$^{-1}$. It is phase delayed by 60 degrees. Note that in X-band
because of the AFC only the absorptive part of the ESR is measured. For Q-band
with $P$=10 mW, $\lvert E_1 \rvert_{max}$ is $\sim$200 V/m and $\Delta$ {\bf k} becomes
0.5$\times 10^5$ m$^{-1}$. Because of the $v\times B_0$-term, it is much smaller and its phase
is -70 degrees with respect to the $B_1$-field. The ESR of the 2D electrons should thus be
almost completely dispersive, which is indeed observed in the experiment (see fig.~\ref{qbandsignal}).
\par
To estimate {\bf B}$_{BR}$ we have to estimate $\alpha_{BR}$ for AlAs, as it is not known.
We note however, that the deviation from the free electron $g$-factor is caused by
SO-interaction and that this deviation for our AlAs is nearly 8 times larger than that
for Si~\cite{Wilamowski2}. For our estimate, we assume $\alpha_{BR}$ to scale with the
SO-interaction, and use the measured $\alpha_{BR}$ of Si (5.5 10$^{-15}$ eVm)~\cite{Wilamowski2}
to calculate {\bf B}$_{BR}$. This effective magnetic field then becomes 14 Gauss in X-band,
almost two orders of magnitude larger than the $B_1$-field (0.2 Gauss). We note that the
estimated $B_{BR}$ is twice larger than the smallest linewidth measured, but given the crudeness
of the model, it agrees surprisingly well with the necessary field strength needed to detect
ESR of 2$\times$ 10$^{10}$ spins. In Q-band both the Dresselhaus and the Bychkov-Rashba
interactions contribute to the effective field and although the estimated $\Delta$ {\bf k}
is much smaller, the Dresselhaus part (that in III-IV semiconductors contains an additional
{\bf k}$^3$ term next to the linear term) could still cause a sizable effective magnetic field.

\begin{figure}[t]
\centering
\includegraphics[width=85mm]{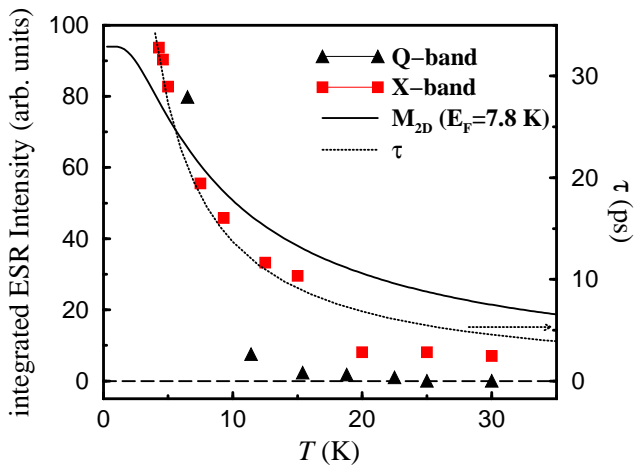}
\caption{Integrated ESR intensity in Q-band ($\blacktriangle$) and X-band ($\blacksquare$)
vs. temperature; $P$=20 mW. Solid line: magnetisation of a 2DEG with
a density of 2.5 10$^{15}$ m$^{-2}$ with two valleys occupied and a mass of 0.46 m$_e$.
($E_F$=7.8~K). Dotted line: the scattering time derived from the measured mobility.
}
\label{tdep}
\end{figure}
\par
Since the scattering time is somewhat shorter than the inverse microwave frequency, only part
of the electrons will be able to follow the $E_1$-field. As the temperature is increased this
part will decrease since the scattering time decreases. Consequently, the temperature dependence
should be stronger than the 2D magnetisation. This is indeed the case.  Fig.~\ref{tdep} plots
the ESR intensity vs. temperature for both X-band ($\blacksquare$) and Q-band ($\blacktriangle$).
The solid line is the 2D magnetisation~\cite{Nestle} for our AlAs 2DEG ($E_F$=7.8~K). This
temperature dependence is clearly much too weak to describe the data. This is not surprising
since the measured scattering time 
is also a rather strong function of temperature (dotted line) and according to the above,
the ESR should represent both the scattering time and the 2D magnetisation. 

\begin{figure}[t]
\centering
\includegraphics[width=85mm]{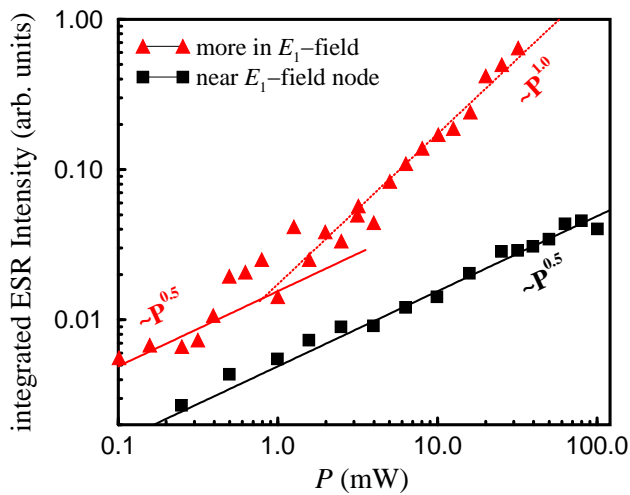}
\caption{Power dependence of the integrated ESR intensity in Q-Band at 28~K. Squares are
measured with the sample close to a node in the $E_1$-field and show an
approximate $\sqrt{P}$ dependence (solid line); triangles are measured with the sample in
more $E_1$-field. For high $P$, the ESR is approximately linear in $P$ (dotted line).
}
\label{pdep}
\end{figure}

\par
To determine the spin dephasing time from the saturation at higher microwave power,
we conducted power dependent measurements. Fig.~\ref{pdep} plots the ESR intensity vs. $P$
in Q-band. The squares were taken with the sample positioned close to a node in the
$E_1$-field and follow a $\sqrt{P}$-dependence, commonly
observed in ESR. It implies that the ESR intensity is proportional to the (in our case effective)
microwave magnetic field. When the sample is positioned more in the $E_1$-field ($\blacktriangle$)
the power dependence changes significantly. For low $P$, the ESR intensity is still approximately
proportional to $\sqrt{P}$, but for higher $P$, instead of saturating, it becomes
approximately linear in $P$.
At high $P$, we envision that the microwaves cause so many spin transitions that the 2D
magnetisation is forced out of thermal equilibrium. If the conductivity of the 2D electrons
($\sigma_{2D}$) for spin up differs from that for spin down, then the power dissipated by the
2D electrons will change at the resonance ($\Delta P \propto \Delta \sigma_{2D} E_1^2$).
To go from a $\sqrt{P}$ to a linear dependence, then implies that
$\Delta \sigma_{2D}$ is proportional to $E_1^2$. Such behaviour is indeed observed in
conductivity measurements on Si/SiGe 2DEGs under microwave radiation~\cite{Graeff}.
\par
In conclusion we have presented direct ESR on a single 2DEG in the III-V semiconductor AlAs.
At 9.35 GHz and 34 GHz $g$-factors of 1.991 and 1.989 were determined. We demonstrated that
the ESR originates from the microwave $E_1$-field as opposed to conventional ESR that
relies on the $B_1$-field. The ESR is attributed to a periodic modulation of the electron wave
vector around $k_F$ due to the $E_1$-field and the high mobility of the 2DEG. Through spin-orbit
interaction, this produces an effective magnetic field that induces the observed spin transitions.
Consequently, the temperature dependence of the ESR is much stronger that the 2D magnetisation,
as it additionally incorporates the temperature dependence of the scattering time of the 2D
electrons.
\par
We thank H.~K\"ummerer and A.~H\"ubel for experimental support and discussions and K.
von Klitzing and G.~Constantini for critical reading of the manuscript.
The work was supported by the DFG, the BMBF (01BM913/0) and the Graduiertenkolleg "Moderne
Methoden der magnetischen Resonanz in der Materialforschung".

\end{document}